\pdfoutput=1
\documentclass[aps,prc,superscriptaddress,nofootinbib,a4paper, twocolumn]{revtex4}
\usepackage{color}
\usepackage{graphicx}
\usepackage{amssymb}
\usepackage[version=3]{mhchem}
\usepackage{ulem}
\begin{document}
\title{$\Lambda\Lambda$ Interaction in a Nuclear Density Functional Theory
and Hyperon Puzzle of the Neutron Star}
\author{Soonchul Choi}
\affiliation{Center for Exotic Nuclear Studies, Institute for Basic Science, Daejeon 34126, Korea}
\author{Emiko Hiyama}
\affiliation{Department of Physics, Tohoku University, Sendai 980-8578, Japan}
\affiliation{Nishina Center, RIKEN, Wako, Saitama 351-0198, Japan}
\author{Chang Ho Hyun}
\affiliation{Department of Physics Education, Daegu University, Gyeongsan 38453, Korea}
\author{Myung-Ki Cheoun}
\affiliation{Department of Physics and OMEG Institute, Soongsil University, Seoul 06978, Korea}

\date{\today}

\begin{abstract}
A Skyrme-type effective potential is determined to describe the interaction between $\Lambda$ hyperons in nuclear medium.
Experimental data of the binding energies of the double-$\Lambda$ ($\Lambda\Lambda$) nuclei with mass numbers $A=10$--$13$ are used 
to fit the parameters of the $\Lambda\Lambda$ interaction.
As a result of the fitting, we obtain eight different sets of the $\Lambda\Lambda$ interaction parameters,
which reproduces the input data within 5\% deviation from the experimental data on average.
The eight $\Lambda\Lambda$ interactions are plugged in the calculation of the heavier $\Lambda\Lambda$ nuclei and the neutron star equation of state
to explore the issue of hyperon puzzle.
We found that the $\Lambda\Lambda$ interaction, specifically, p-wave interaction makes the equation of state stiff enough that the maximum mass
of the neutron star can be as large as, or above $2\;M_\odot$.
\end{abstract}
\maketitle

\section{Introduction}

Unified description of finite nuclei and infinite nuclear matter is a fundamental challenge in the nuclear physics.
Such a model or method, if exists, will cover vast range of physical systems, e.g. several thousands of stable and unstable
nuclei from light to super-heavy elements in the nuclear chart as well as nuclear matter like neutron stars and their thermal nuclear reactions in the stellar evolution.
The unification scheme can also be extended to the systems that have non-zero expectation values of the strangeness.
Density functional theory is a promising candidate that can provide a framework for the principle-based unified model for the nuclear 
many-body systems.
However, calculation of the nuclear energy density functional (EDF) from the first principles is still a challenging issue,
and it is non-trivial to estimate when such an approach can provide quantitative description and prediction for both the
finite nuclei and infinite nuclear matter.
A way around is needed.
Inspired by the idea of the low-energy effective field theory, it was suggested to expand the nuclear EDF
in the power of $\rho^{1/3}$ where $\rho$ is the baryon density,
and determine unknown constants with experimental or empirical data of nuclear matter and finite nuclei 
\cite{kidsnm, kids-acta, kids-npsm2017, kids-nuclei1, kids-nuclei2}.
The expansion scheme is named KIDS (Korea-IBS-Daegu-SKKU) density functional framework.

In a recent work \cite{hypernuclei1}, we performed a first application of the KIDS formalism to the interaction of strangeness in nuclear medium.
With the multiple density-dependent terms for the many-body interaction of $\Lambda$ hyperons,
experimental data of single-$\Lambda$ hypernuclei are reproduced with high accuracy.
We have examined the perturbative behavior of the expansion,
and verified that the result shows convergent behavior as the higher order terms are added. 
Interestingly, threshold creation density of the $\Lambda$ hyperon in the neutron star shows tendency significantly different from the results in the literature,
and we found that the nuclear symmetry energy is a critical source of the difference.

One big issue of the neutron star physics is to reduce the uncertainty in the equation of state (EoS) at densities above the 
nuclear saturation density ($\rho_0$).
In the density region, if $\Lambda$ hyperons are created, generally it makes the EoS soft,
so their presence is not compatible with the observation of neutron star masses larger than $2M_\odot$ \cite{Demorest:2010bx,Antoniadis:2013pzd,NANOGrav:2019jur}.
However, it is shown that inclusion of the interactions between $\Lambda$ hyperons can make the EoS
stiff enough to support $2M_\odot$ mass even if $\Lambda$ hyperons exist in the neutron star \cite{ijmpe2015}. 

Recently, 
double $\Lambda$-Be isotopes have been observed, although we have some
interpretations to identify binding energies and states \cite{be1011,be12,b1213}. 
By these observations, we obtained information on $s-$wave term. 
However,  to draw a definite conclusion about the role of $\Lambda$ hyperons in the neutron star, it is necessary to obtain more information on $\Lambda \Lambda$ interaction such as $p-$ wave term and many-body terms.

In the present work, by making use of the available data on the binding energies of $\Lambda\Lambda$ nuclei,
we determine the parameters of $\Lambda\Lambda$ interaction in the non-relativistic Skyrme-type effective potential.
In the former versions of Skyrme-type $\Lambda\Lambda$ interaction, because of the limitation in the available data,
only two parameters are determined, and the remaining two are assumed to be zero \cite{lans1998, mina2011}, 
but now it becomes feasible to determine the four parameters.
We fit the four parameters to the binding energies of $\Lambda\Lambda$ nuclei in the mass range $A=10$--$13$ \cite{Danysz,be1011,be12, b1213,Nakazawa}.
We account for the uncertainty of the interaction as broad as possible, 
and as a result eight sets of parameters for the $\Lambda\Lambda$ interaction are obtained. 

With the new parameter sets for the $\Lambda\Lambda$ interaction,
we solve the charge neutrality and $\beta$ equilibrium equations at a given baryon density and determine the EoS of the neutron star matter.
Integrating the Tolman-Oppenheimer-Volkoff (TOV) equations, we obtain the mass-radius relation of the neutron star for the eight different $\Lambda\Lambda$ interactions.

When there are only nucleons, the models satisfy the constraints $R_{1.4} = 11.8$--$13.1$ km obtained by the NICER analysis \cite{Raaijmakers:2019qny} and  the large maximum mass condition $M_{\rm max} \gtrsim 2.0 M_\odot$, where $R_{1.4}$ denotes the radius of stars with mass $1.4 M_\odot$.
If we include the $N\Lambda$ interaction, the EoS is softened substantially that the maximum mass does not exceed $1.9 M_\odot$.
By including the $\Lambda\Lambda$ interaction, the EoS of the neutron star matter becomes stiff again.
The stiffness depends on the parameter set, so we have EoSs ranging from softer to as stiff as the nucleon matter.
In detail, two sets give EoS slightly softer than that only by the nucleon matter,
and four sets are similar to the nucleon matter.
\footnote{For the two other parameter sets, convergent solutions are not obtained for the neutron star matter EoS.}
As a consequence, four sets satisfy the condition $M_{\rm max} \geq 2.0 M_\odot$,
and two soft sets give $M_{\rm max}$ slightly smaller than $2M_\odot$.

The ambiguity surrounding the character of soft and stiff sets comes from light hypernuclei and the behavior of these interactions in nuclear matter remain.
To address these questions and deepen our understanding, we turn our attention to heavier $\Lambda\Lambda$ hypernuclei, $^{18}_{\Lambda\Lambda}$O, 
$^{42}_{\Lambda\Lambda}$Ca, $^{92}_{\Lambda\Lambda}$Zr, and $^{210}_{\Lambda\Lambda}$Pb.
In particular, the potential to distinguish the complex many-body interaction emerges notably within the binding energies of $^{42}_{\Lambda\Lambda}$Ca and $^{92}_{\Lambda\Lambda}$Zr nuclei.

The paper is organized as follows.
In Sec. II, we introduce the model, and discuss the result of $\Lambda\Lambda$ interaction determined by fitting to
$\Lambda\Lambda$ hypernuclei data.
Section III presents the result for neutron stars.
We summarize the work in Sec. IV.

\section{$\Lambda\Lambda$ interaction}

We start with a Hamiltonian density $H = H_N + H_\Lambda + H_{\Lambda\Lambda}$,
where $H_N$ is the pure nucleon part ($NN$ interaction), $H_\Lambda$ consists of the single-$\Lambda$ mean field 
terms ($N\Lambda$ interaction), and $\Lambda\Lambda$ interactions are included in $H_{\Lambda\Lambda}$.
Nucleon part is given in the KIDS framework as
\begin{eqnarray}
H_N &=& \frac{\hbar^2}{2m_N} \tau_N + \frac{3}{8}t_0 \rho^2_N - \frac{1}{8}t_0(1+2x_0) \rho^2_N \delta^2 \nonumber \\ 
&& + \frac{1}{16} \sum_{n=1}^3 t_{3n} \rho^{2+n/3}_N
- \frac{1}{48} \sum_{n=1}^4 t_{3n}(1+2x_{3n}) \rho^{2+n/3}_N \delta^2 \nonumber \\ 
&& + \frac{1}{64} (9t_1 - 5t_2 -4 t_2 x_2) (\nabla \rho_N)^2 \nonumber \\
&& - \frac{1}{64}(3t_1 + 6t_1 x_1 +t_2 + 2 t_2 x_2) (\nabla \rho_N \delta)^2 \nonumber \\
&& + \frac{1}{8} (2t_1 + t_1 x_1 + 2t_2 + t_2 x_2) \tau_N \rho_N \nonumber \\
&& -\frac{1}{8} (t_1 + t_1 x_1 - t_2 -2 t_2 x_2) \sum\rho_q\tau_q \nonumber \\
&& + \frac{1}{2} W_0 \left( \nabla \rho_N \cdot \mathbf{J}_N + \sum_q \nabla \rho_q \cdot \mathbf{J}_q \right),
\end{eqnarray}
where $\tau_N$ denotes the kinetic energy density of the nucleon, $\delta = (\rho_n - \rho_p)/(\rho_n + \rho_p)$, and $q=n$, $p$.
Single-$\Lambda$ Hamiltonian density reads
\begin{eqnarray}
H_\Lambda &=& \frac{\hbar^2}{2m_\Lambda} \tau_\Lambda + u_0 \left(1+\frac{1}{2} y_0 \right) \rho_N \rho_\Lambda \nonumber \\
&& + \frac{3}{8} \rho_\Lambda \sum_{n=1}^4 u_{3n} \left(1+\frac{1}{2}y_{3n}\right) \rho^{1+n/3}_N \nonumber \\
&& + \frac{1}{4}(u_1+u_2)(\tau_\Lambda \rho_N + \tau_N \rho_\Lambda) \nonumber \\ 
&& + \frac{1}{8}(3u_1 - u_2)(\nabla\rho_N \cdot \nabla\rho_\Lambda),
\end{eqnarray}
and $\Lambda\Lambda$ interaction is given as
\begin{eqnarray}
H_{\Lambda\Lambda} &=& 
\frac{1}{4} a_0 \rho^2_\Lambda + \frac{1}{8} (a_1 +3a_2)\rho_\Lambda \tau_\Lambda \nonumber \\
&& + \frac{3}{32}(a_2 - a_1) \rho_\Lambda \nabla^2 \rho_\Lambda + \frac{1}{4} a_3 \rho^2_\Lambda \rho_N,
\label{eq:hll}
\end{eqnarray}
where the $a_0$, $a_1$, $a_2$, and $a_3$ are parameters.
Here we note that the $a_0$ and $a_1$ terms and the $a_2$ term correspond to the contributions by s-wave and p-wave interactions, respectively \cite{Vautherin:1971aw}.

To determine the parameters using fitting, we need more than four $\Lambda\Lambda$ hypernuclei data.
Currently, we have reliable four double $\Lambda$ hypernuclei, $\ce{^6_{\Lambda\Lambda}}\rm{He}$ \cite{Takahashi:2001nm}, $\ce{^{10}_{\Lambda\Lambda}}\rm{Be}$ \cite{Danysz}, $\ce{^11_{\Lambda\Lambda}}\rm{Be}$ \cite{be1011}, and $\ce{^13_{\Lambda\Lambda}}\rm{B}$ \cite{Nakazawa}.
Among them, $\ce{^6_{\Lambda\Lambda}}\rm{He}$ is too light to fit the parameter within our framework.
Thus, we use the data of $\ce{^{10}_{\Lambda\Lambda}}\rm{Be}$, $\ce{^{11}_{\Lambda\Lambda}}\rm{Be}$, and $\ce{^{13}_{\Lambda\Lambda}\rm{Be}}$.
However, to fix the parameters, we need more data.
Along this line, we decided to use more data such as $\ce{^{12}_{\Lambda\Lambda}}\rm{Be}$, and $\ce{^{12}_{\Lambda\Lambda}}\rm{B}$ in Ref.~\cite{be12, b1213} respectively, although there are ambiguities to identify the $\Lambda\Lambda$ hypernuclei.
The significance lies in the scenario involving $\ce{^{12}_{\Lambda\Lambda}}\rm{Be}$, stemming from a captivating HIDA event. This interpretation unfolds into two compelling possibilities: either as $\ce{^{11}_{\Lambda\Lambda}}\rm{Be}$ or $\ce{^{12}_{\Lambda\Lambda}}\rm{Be}$. Taking into account the distinct outcomes of the existing $\ce{^{11}_{\Lambda\Lambda}}\rm{Be}$ investigations \cite{be1011}, our research adopts a strategic approach.
Although confirming $\ce{^{12}_{\Lambda\Lambda}}\rm{B}$ interpretation~\cite{b1213} necessitates further experimentation, theoretical calculations in Gal's study \cite{Gal:2011zr} do not strongly challenge this prospect. We elect the interpretations of $\ce{^{12}_{\Lambda\Lambda}}\rm{Be}$ and $\ce{^{12}_{\Lambda\Lambda}}\rm{B}$ to enrich parameter determination and refine model.



In \cite{hypernuclei1}, we explored the effect of symmetry energy to the creation of $\Lambda$ hyperons within the neutron star.
We find that the dependence on the symmetry energy is critical, so the creation density varies widely from 0.5 fm$^{-3}$ to 0.7 fm$^{-3}$.
In order to take into account the uncertainty of the creation density maximally, we choose two models KIDS0 and KIDS-D 
that give the largest (0.7 fm$^{-3}$) and the smallest (0.5 fm$^{-3}$) $\Lambda$-hyperon creation density, respectively.
Numerical values of the parameters for the KIDS0 and KIDS-D models can be found in \cite{kids-nuclei2} and \cite{npsm2021}, respectively,
and those including the single-$\Lambda$ interactions are given in \cite{hypernuclei1}.

The $\Lambda\Lambda$ interaction given by Eq.~(\ref{eq:hll}) have been employed in the precedent Skyrme force works \cite{lans1998, mina2011}.
In these works, due to the lack of available data, $a_0$ and $a_1$ are determined from the fitting, but $a_2$ and $a_3$ are assumed to be zero.
In other words, they did not include p-wave and many-body interactions.
As a consequence, the interaction is constrained poorly, and it shows wide range of stiffness.
More data for the double-$\Lambda$ hypernuclei have been reported afterwards,
so now it is feasible to determine all the four parameters $a_0$--$a_3$ without assuming zero values for any of them. In the fitting we presume the sign of $a_0$ negative.
The $a_0$ term represents the two-body $\Lambda\Lambda$ interaction.
It is shown that this force, when it is fitted to Nijmegen two-body potential, is attractive around the nuclear saturation density \cite{prc2002}.
In order to classify the parameter sets, we perform the fitting for both positive and negative signs of $a_1$, $a_2$ and $a_3$,
which gives eight sets of parameters for each KIDS0 and KIDS-D model.
However, for certain combinations of the sign, input data are not reproduced accurately.
Discarding those relatively worse fitting result, we obtain four sets of the parameter for each KIDS0 and KIDS-D model
as shown in Tab.~\ref{tab1}.
As mentioned already, other combinations of the sign give MD larger than 5\%, so they are excluded from the consideration.
As a whole, signs and magnitudes of the parameters in KIDS0 are similar to those by KIDS-D model.
An interesting feature is when $a_3$ is positive (LL2 and LL4), the magnitudes are smaller than those of other parameters by orders of magnitude,
but such a suppression is not seen when $a_3$ is negative.
Physical meaning of the origin of the small density-dependent term is $\Lambda\Lambda$--$\Xi N$ coupling \cite{ptp1994}.
The $\Lambda\Lambda$--$\Xi N$ coupling is strongly suppressed in heavy nuclei due to the Pauli blocking effect, 
so the small magnitude of $a_3$ is consistent with the suppression mechanism.
In addition, the order of magnitude of $a_0$ in the LL4 sets are in better agreement with the values 
obtained from the fitting in \cite{lans1998, mina2011} which are independent of this work.

\begin{table}
    \begin{center}
    \begin{tabular}{c|cccc}\hline 
             \phantom{aa}& \phantom{aa} $a_0$ \phantom{aa} & \phantom{aa} $a_1$ \phantom{aa} & \phantom{aa} $a_2$ \phantom{aa} & \phantom{aa} $a_3$ \phantom{aa} \\ \hline
             KIDS0-LL1 & $-9.522$  & 1112.782 & 13684.952 & $-5226.032$  \\
             KIDS0-LL2 & $-2804.779$ & 7127.180 & $-1710.144$ & 1.970  \\
             KIDS0-LL3 & $-1809.005$ & 8587.169 & $-2352.189$ & $-12159.633$   \\
             KIDS0-LL4 & $-144.843$  & $-44.429$  & 4141.366  & 36.590  \\
             \hline
             KIDS-D-LL1 & $-222.557$  & 1641.863  & 14826.149  & $-5771.724$  \\
             KIDS-D-LL2 & $-2700.449$ & 6669.803  & $-1668.553$  & 19.523 \\    
             KIDS-D-LL3 & $-2846.745$ & 12571.015 & $-2745.506$ & $-13396.543$  \\
             KIDS-D-LL4 & $-194.167$  & $-16.981$   & 3488.147   & 0.690 \\
             \hline
    \end{tabular}
    \end{center}
    \caption{Parameters for the $\Lambda\Lambda$ interaction.
    Units of parameters are MeV~fm$^{3}$ for $a_0$, MeV~fm$^{5}$ for $a_1$ and $a_2$, and MeV~fm$^6$ for $a_3$.
    }
    \label{tab1}
\end{table}


In this work, we use the binding energies of five double-$\Lambda$ hypernuclear data tabulated in Tab.~\ref{tab2}.
The table also shows the binding energy of the theory obtained as a result of the fitting.
Accuracy of the fitting is measured by the mean deviation (MD)
\begin{equation}
{\rm MD} = \frac{1}{N_d} \sum^{N_d}_{i=1} \left| \frac{E^{\rm exp}_i - E^{\rm calc}_i}{E^{\rm exp}_i} \right|\times 100,
\end{equation}
where $N_d$ denotes the number of experimental data, and we use the center values of data for $E^{\rm exp}_i$.
In the last column of Tab.~\ref{tab2}, MD values are shown.
Because of the insufficient number of data, the accuracy of the agreement is not as good as the single-$\Lambda$ hyperon interactions \cite{hypernuclei1}.

\begin{table*}
        \begin{center}
        \begin{tabular}{c|cccccc}\hline
         & $\ce{^{10}_{\Lambda\Lambda}}{\rm Be}$ \cite{Danysz} & $\ce{^{11}_{\Lambda\Lambda}}{\rm Be}$ \cite{be1011} 
         & $\ce{^{12}_{\Lambda\Lambda}}{\rm Be}$ \cite{be12} 
         & $\ce{^{12}_{\Lambda\Lambda}}{\rm B}$ \cite{b1213} 
         & $\ce{^{13}_{\Lambda\Lambda}}{\rm B}$ \cite{Nakazawa}
         & MD (\%) \\ \hline
        Exp.  &\phantom{a} 15.14 \phantom{a} & 19.07$\pm$0.11 \phantom{a} & 22.23$\pm$1.15 \phantom{a}
    & 20.60$\pm$0.74 \phantom{a}  & 23.30$\pm$0.70 \\
        KIDS0-LL1 & 15.370 & 18.608 & 21.147 & 21.794 & 23.395 & 3.338 \\
        KIDS0-LL2 & 15.065 & 18.573 & 20.979 & 21.557 & 22.889 & 4.474 \\
        KIDS0-LL3 & 15.195 & 18.895 & 21.440 & 22.129 & 23.384 & 2.857 \\
        KIDS0-LL4 & 15.786 & 18.769 & 21.124 & 21.753 & 23.000 & 3.876 \\
        \hline
        KIDS-D-LL1 & 15.425 & 18.558 & 21.209 & 21.882 & 23.558 & 3.632 \\
        KIDS-D-LL2 & 15.878 & 18.405 & 20.828 & 21.399 & 22.761 & 4.509 \\    
        KIDS-D-LL3 & 15.171 & 18.793 & 21.417 & 22.065 & 23.762 & 3.217 \\
        KIDS-D-LL4 & 15.862 & 18.466 & 21.043 & 21.625 & 22.952 & 4.285 \\
        \hline
        \end{tabular}
        \end{center}
        \caption{Binding energy of the $\Lambda\Lambda$ hypernuclei. 
        Shown are the experimental data \cite{Danysz, be1011,be12, b1213, Nakazawa} used in the fitting, and the result of the reproduced values. 
        Binding energies are in MeV.
        In the last column, mean deviation from experimental data are shown in the unit of \%.}
        \label{tab2}
\end{table*}

\begin{figure*}
\begin{center}
\includegraphics[width=0.4\textwidth]{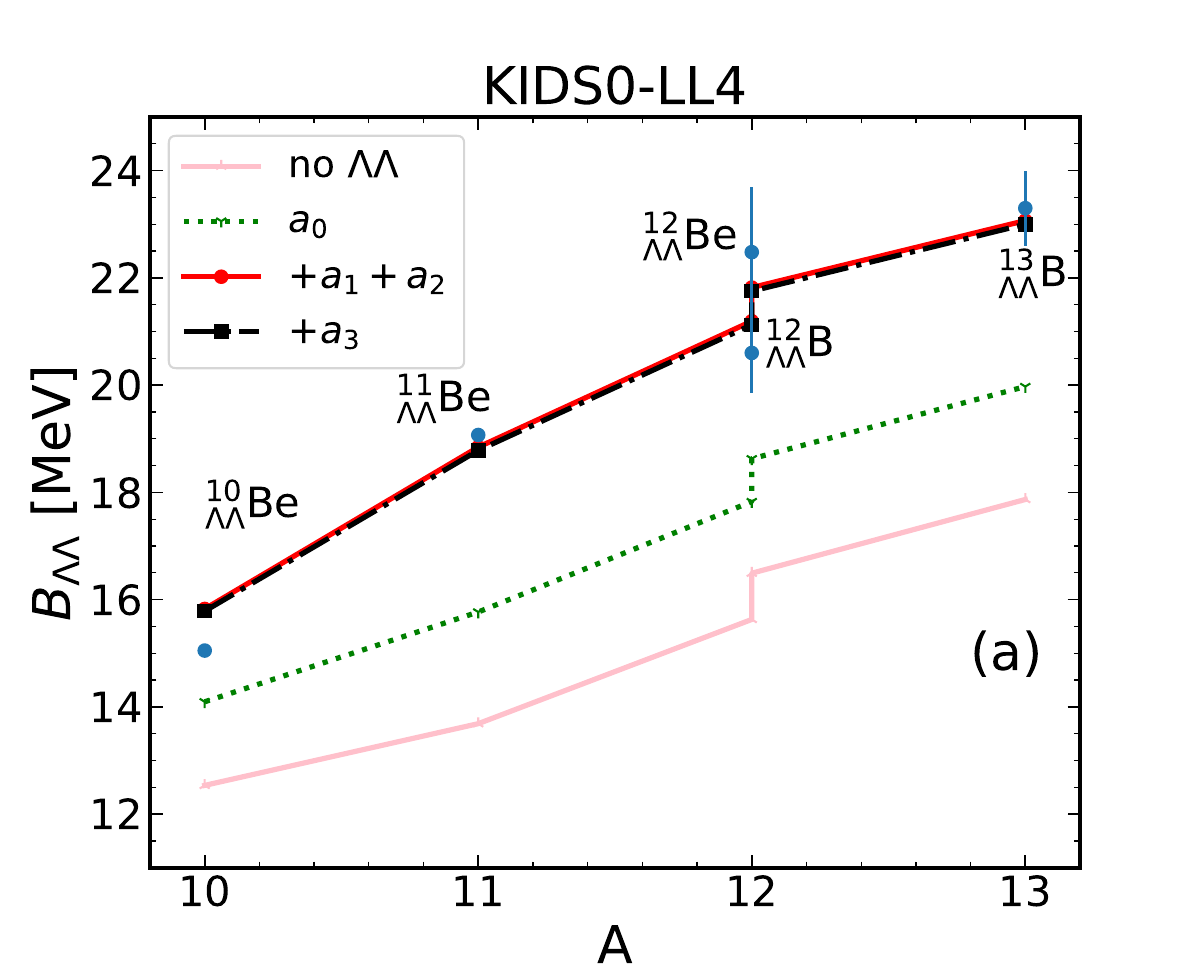}
\includegraphics[width=0.4\textwidth]{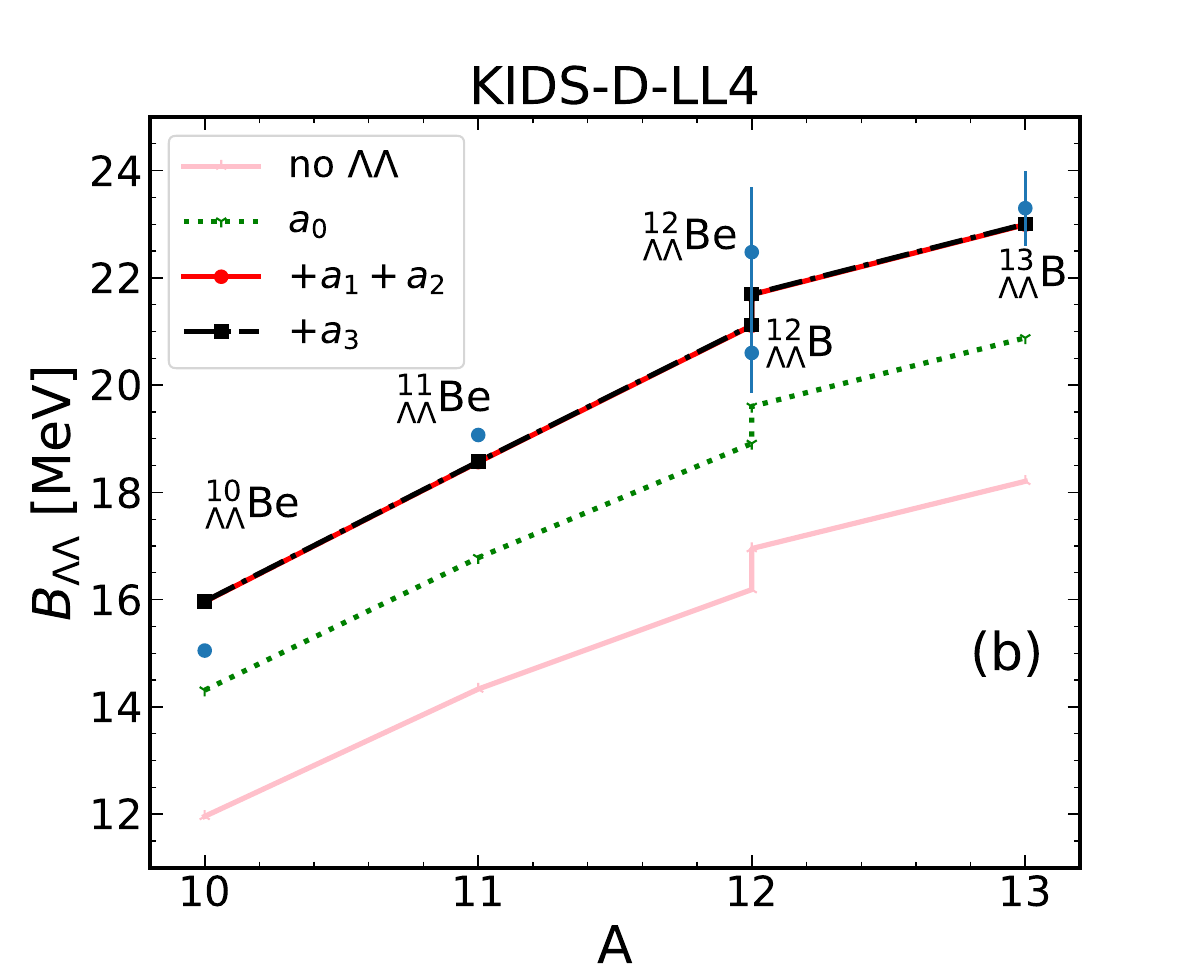}
\end{center}
\caption{Disentanglement of contributions by each term for (a) KIDS0-LL4 and (b) KIDS-D-LL4.
Solid-magenta line denotes the binding energy without $\Lambda\Lambda$ interaction, and dot-green line
shows the result by adding $a_0$ term to the `no$\Lambda\Lambda$' result.
Solid-red line is for the result adding momentum-dependent terms to the dot-green line,
and the dot-dash-black line presents the full result.}
\label{fig1}
\end{figure*}

Taking these parameter sets into consideration,
we plot the contribution of each term in the $\Lambda\Lambda$ interaction
to the binding energies for the KIDS0-LL4 and KIDS-D-LL4 models in Fig.~\ref{fig1}.

Solid-magenta lines show the $\Lambda\Lambda$ binding energies ($B_{\Lambda\Lambda}=E(\ce{^A_{\Lambda\Lambda}Z})-E(\ce{^{A-2} Z})$) without $\Lambda\Lambda$ interaction, but with only $NN$ and $N\Lambda$ interactions.
Proportion of $NN$ and $N\Lambda$ interactions takes about 75--80\% of the total binding energy.
Thus, $NN$ and $N\Lambda$ interactions are most basic and dominant in the binding energy,
and at the same time, $\Lambda\Lambda$ interaction plays a crucial role to obtain quantitative agreement with experiment.
Dot-green lines correspond to the result in which the contribution of $a_0$ term in the $\Lambda\Lambda$ interaction is added to the result of $NN$ and $N\Lambda$ 
contributions (`no $\Lambda\Lambda$'), and solid-red lines show the result of $NN + N\Lambda + \Lambda\Lambda$.
Half of the difference between `no $\Lambda\Lambda$' and experiment is compensated by attractive $a_0$ term,
and the remaining half is filled by the momentum-dependent terms $a_1$ and $a_2$.
Specifically, the competition of p- and s-wave term by the ($a_2-a_1$) strength plays a significant attractive force role on light $\Lambda\Lambda$ nuclei because of the negative $\nabla \rho$ in the $\nabla^2 \rho$ term factor.

The result with the density-dependent term $a_3$ representing many-body interactions including hyperon
is depicted by the dot-dash-black lines.
Correction by the $a_3$ terms is very marginal.
Small contribution of the density-dependent term in the $\Lambda\Lambda$ interaction can be explained
in terms of small magnitude of $a_3$ compared to other parameters.
Suppression of $a_3$ term is contradictory to the $NN$ and $N\Lambda$ interactions,
in which the parameters for the density-dependent terms are similar in magnitude to the two-body terms.
It means that the many-body interactions relevant to $\Lambda$ hyperons are smaller than those only by nucleons.
Effect of the density-dependent terms might be more evident at heavier $\Lambda\Lambda$ nuclei as well as high densities, so the neutron star provides a good laboratory to explore the role of each term in the $\Lambda\Lambda$ interaction.

\section{Result}

Results are presented for the neutron star.
In the calculation of the neutron star matter EoS, we encounter a convergence problem for the KIDS0-LL3 and KIDS-D-LL3 models.
Therefore, the results are shown for six models, LL1, LL2 and LL4 parameters with KIDS0 and KIDS-D.

\subsection{Particle Fraction}

\begin{figure*}
\begin{center}
\includegraphics[width=0.4\textwidth]{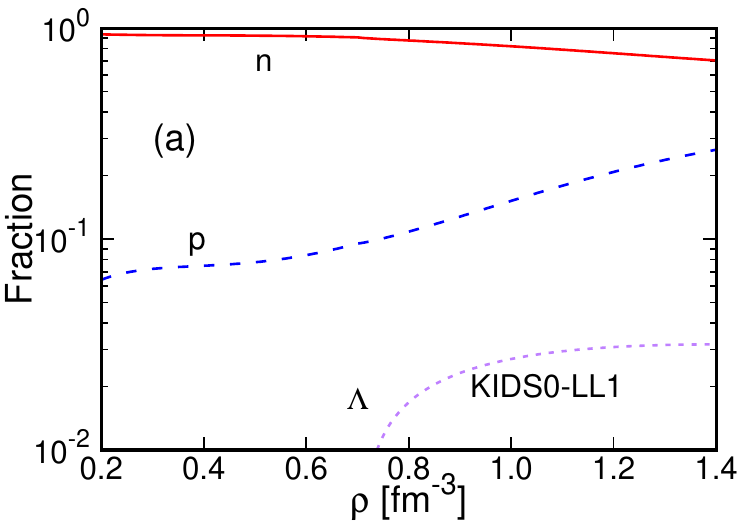}
\includegraphics[width=0.4\textwidth]{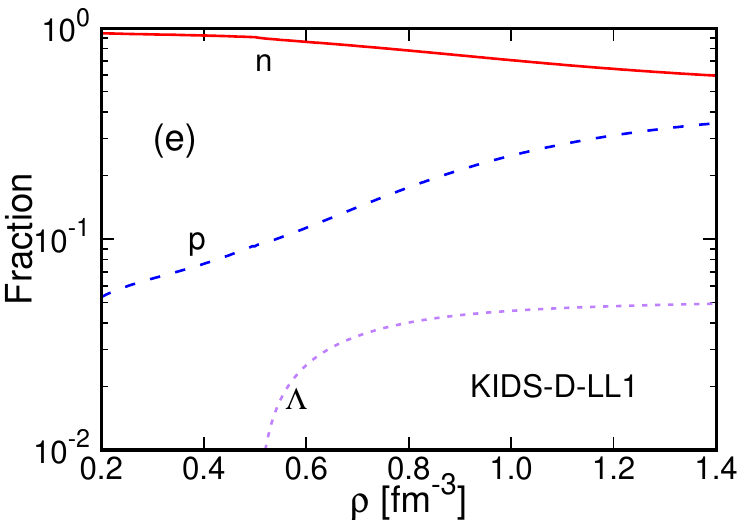}
\includegraphics[width=0.4\textwidth]{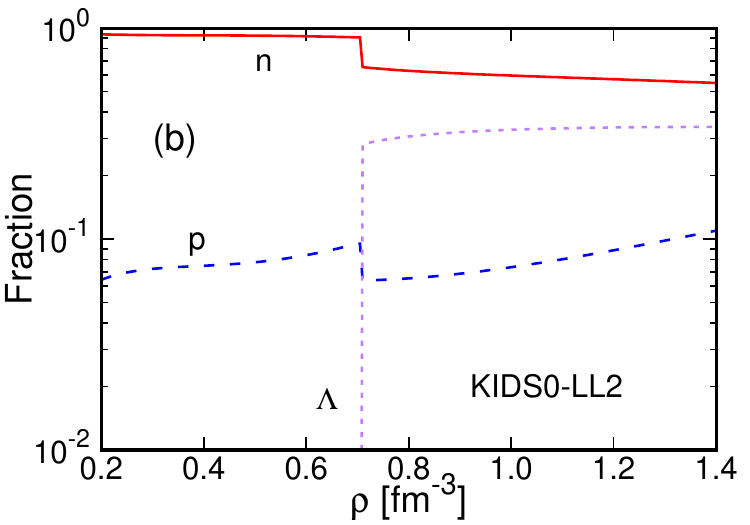}
\includegraphics[width=0.4\textwidth]{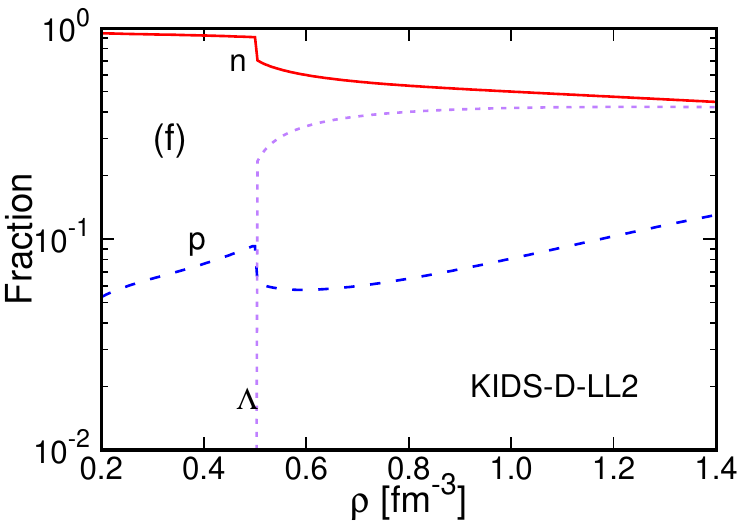}
\includegraphics[width=0.4\textwidth]{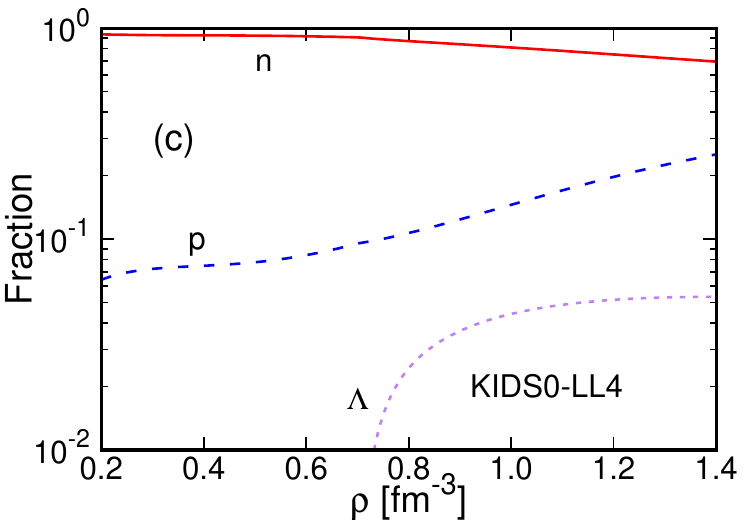}
\includegraphics[width=0.4\textwidth]{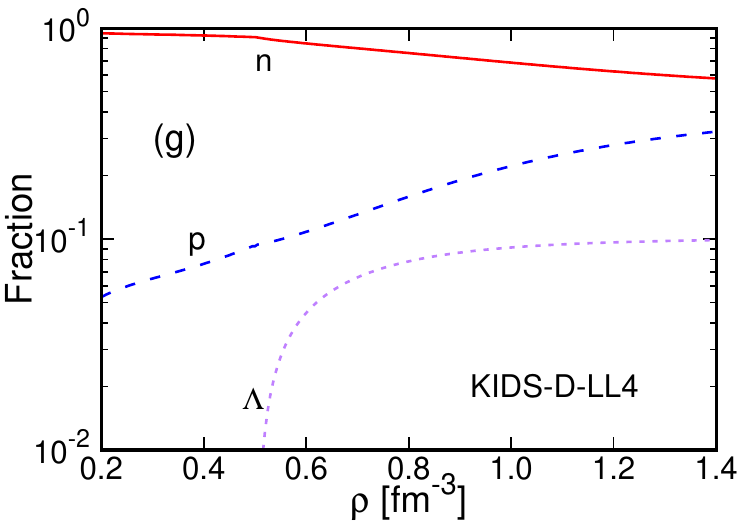}
\includegraphics[width=0.4\textwidth]{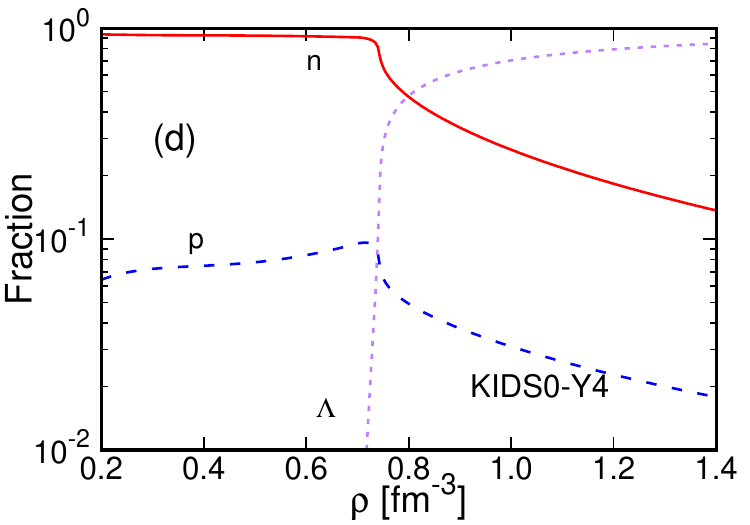}
\includegraphics[width=0.4\textwidth]{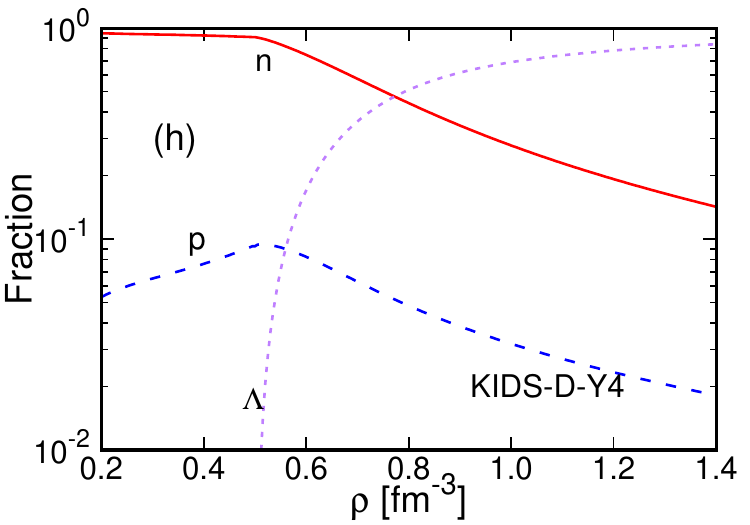}
\end{center}
\caption{Particle fraction of the KIDS0-LL model (left column) and the KIDS-D-LL model (right column).
Panels in the bottom row show the fractions by the single-$\Lambda$ model, KIDS0-Y4 and KIDS-D-Y4, for comparison.}
\label{fig2}
\end{figure*}

\begin{figure*}
    \begin{center}
    \includegraphics[width=0.45\textwidth]{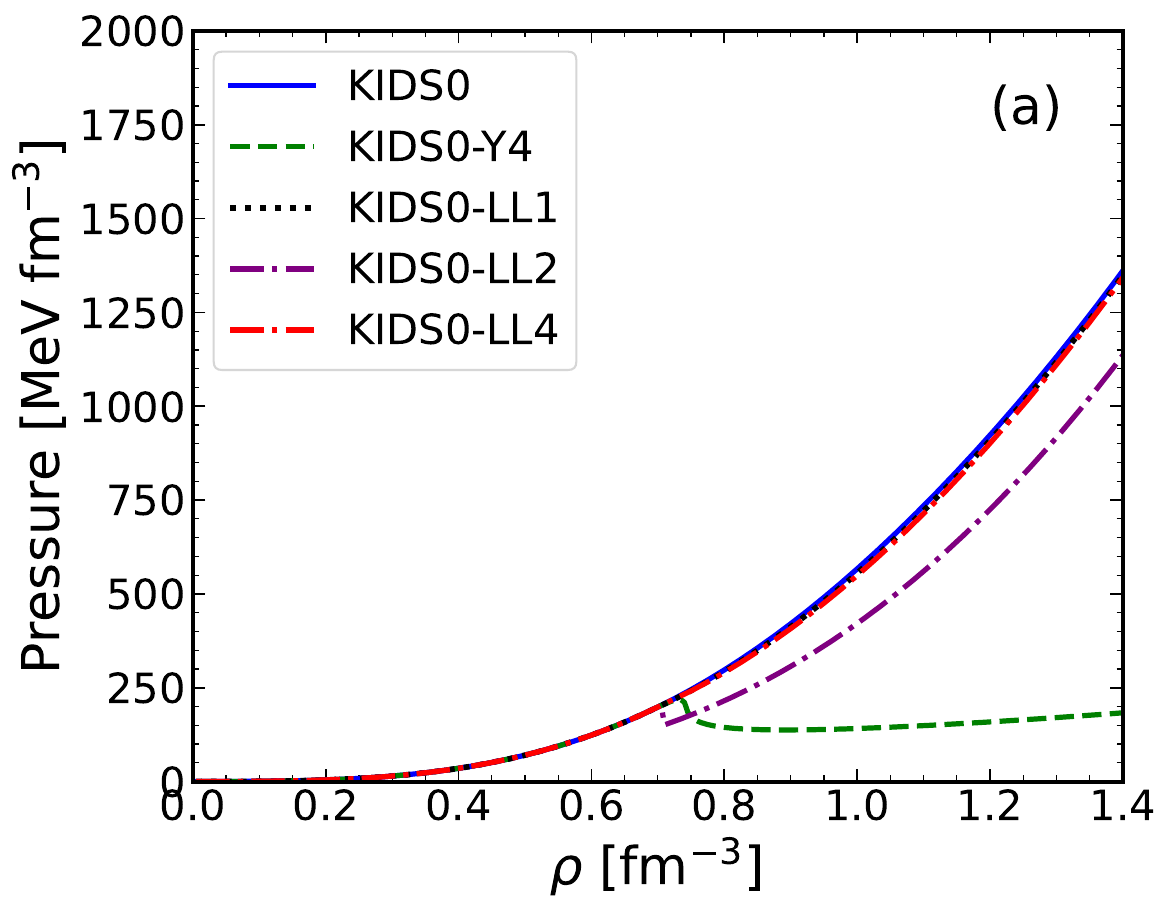}
    \includegraphics[width=0.45\textwidth]{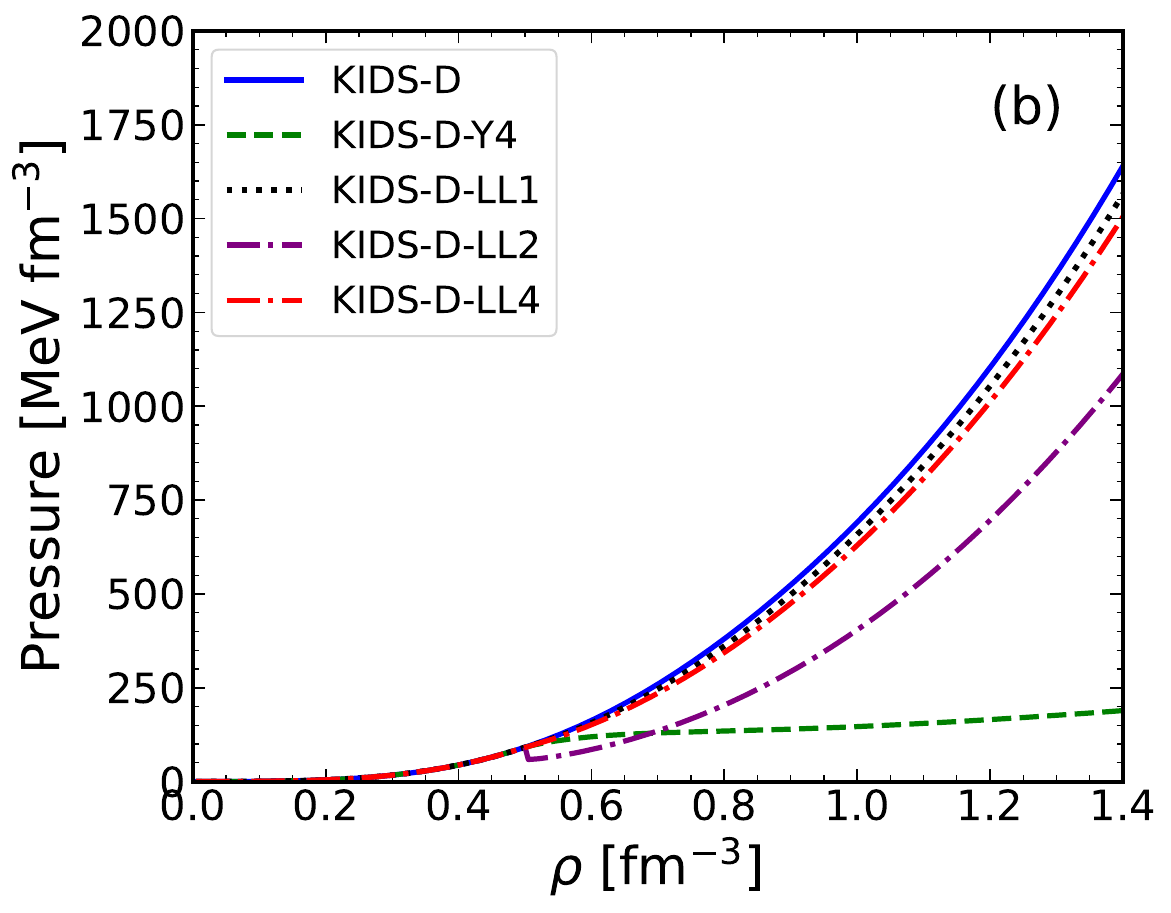}
    \end{center}
    \caption{
        Pressure as a function of baryon density for the (a) KIDS0 and (b) KIDS-D models.
    Lines denoted by KIDS0 and KIDS-D (no $\Lambda$) are from \cite{kids-nuclei2}, and those by KIDS0-Y4 and KIDS-D-Y4 (single $\Lambda$) are from \cite{hypernuclei1}.
    Results with the $\Lambda\Lambda$ interaction by KIDS0-LL and KIDS-D-LL are from the present work.
    }
    \label{fig3}
\end{figure*}

Energy density and pressure of the neutron star core are determined as functions of the density of particles
$n$, $p$, $e$, $\mu$ and $\Lambda$.
Density of each particle is obtained by solving the coupled equations from the $\beta$-equilibirium, charge neutrality,
and the baryon number conservation at a given density $\rho$:
\begin{eqnarray}
\mu_n &=& \mu_p + \mu_e, \,\,\,
\mu_\mu = \mu_e, \,\,\,
\mu_\Lambda = \mu_n, \nonumber \\
\rho_p &=& \rho_e + \rho_\mu, \nonumber \\
\rho &=& \rho_n + \rho_p + \rho_\Lambda.
\end{eqnarray}

Figure~\ref{fig2} shows the result of particle fraction which is defined as the density of a particle divided by
the total baryon density $\rho$.
Panels in the left column exhibit the result of KIDS0-LL models,
and the right column represents the result of the KIDS-D-LL models.
As reported in \cite{hypernuclei1} and presented in the lowermost panel in Fig.~\ref{fig2}, critical (threshold) density $\rho_{\rm crit}$ for the $\Lambda$ creation is 0.7~fm$^{-3}$ 
in KIDS0-Y4 and 0.5~fm$^{-3}$ in KIDS-D-Y4.
In both models, where the $\Lambda\Lambda$ interaction is not considered,
$\Lambda$ hyperon leads to fast decrease of {the neutron fractions.}
Inclusion of the $\Lambda\Lambda$ interaction gives rise to a dramatic change from the Y4 model result.
The fraction of $\Lambda$ is suppressed by orders of magnitude, and it restores the neutron and the proton fractions similar to
the values in which there are only nucleons.
Nevertheless, the detailed result shows sensitivity to both the symmetry energy and the $\Lambda\Lambda$ interaction.

The difference between the KIDS0-LL and KIDS-D-LL models can be understood in terms of the critical density at which
creation of the $\Lambda$ hyperon begins.
In the KIDS-D model, $\Lambda$ hyperons are created at densities much lower than the KIDS0 model,
so it leads to a relatively large fraction of the $\Lambda$ hyperon in the KIDS-D-LL models.

For the LL1 and LL4 parameter sets, $\Lambda$ fraction does not exceed 10\%.
Suppression of the $\Lambda$ hyperon in the LL1 and LL4 sets could be understood by means of the parameters $a_1$ and $a_2$.
In the homogeneous matter, derivative term ($a_2-a_1$) on the r.h.s. of Eq.~(\ref{eq:hll}) vanishes, so the remaining three terms, whose main term is $\tau_\Lambda$, contribute to $\mu_\Lambda$.
With the $a_1$ and $a_2$ values given in Tab.~\ref{tab1}, the second term ($a_1+3a_2$) in the r.h.s. of Eq.~(\ref{eq:hll}) (call it $\rho\tau$ term)
is always positive, {\it i.e.}, repulsive, which prevents significantly larger $\Lambda$ fraction.
For the LL1 set, the magnitude of the coefficient of the $\rho\tau$ term is overwhelming those of $a_0$ and $a_3$.
With a large value of the coefficient, small number of $\rho_\Lambda$ can satisfy the condition of the chemical potential $\mu_\Lambda = \mu_n$,
so the $\Lambda$ fraction becomes small.
Comparing the LL1 parameters with those of LL4,
coefficient of the $\rho\tau$ term is much bigger for LL1, which can explain small $\Lambda$ fraction in LL1 compared to LL4.

For the LL2 parameters, coefficient of the $\rho\tau$ term is much smaller than the LL1 and LL4 parameters.
More importantly, magnitude of $a_0$ is larger by an order of 10 in the LL2 sets than the LL1 and LL4 sets.
Since $a_0$ is negative, large magnitude of $a_0$ demands large $\rho_\Lambda$ values in the condition $\mu_\Lambda = \mu_n$,
so we have lots amount of $\Lambda$ particles in the LL2 parameters.
In brief, repulsive $\Lambda\Lambda$ interaction reduces largely the $\Lambda$ fraction in LL1 and LL4 models for neutron stars.

\subsection{Equation of State}

\begin{figure*}
    \begin{center}
    \includegraphics[width=0.45\textwidth]{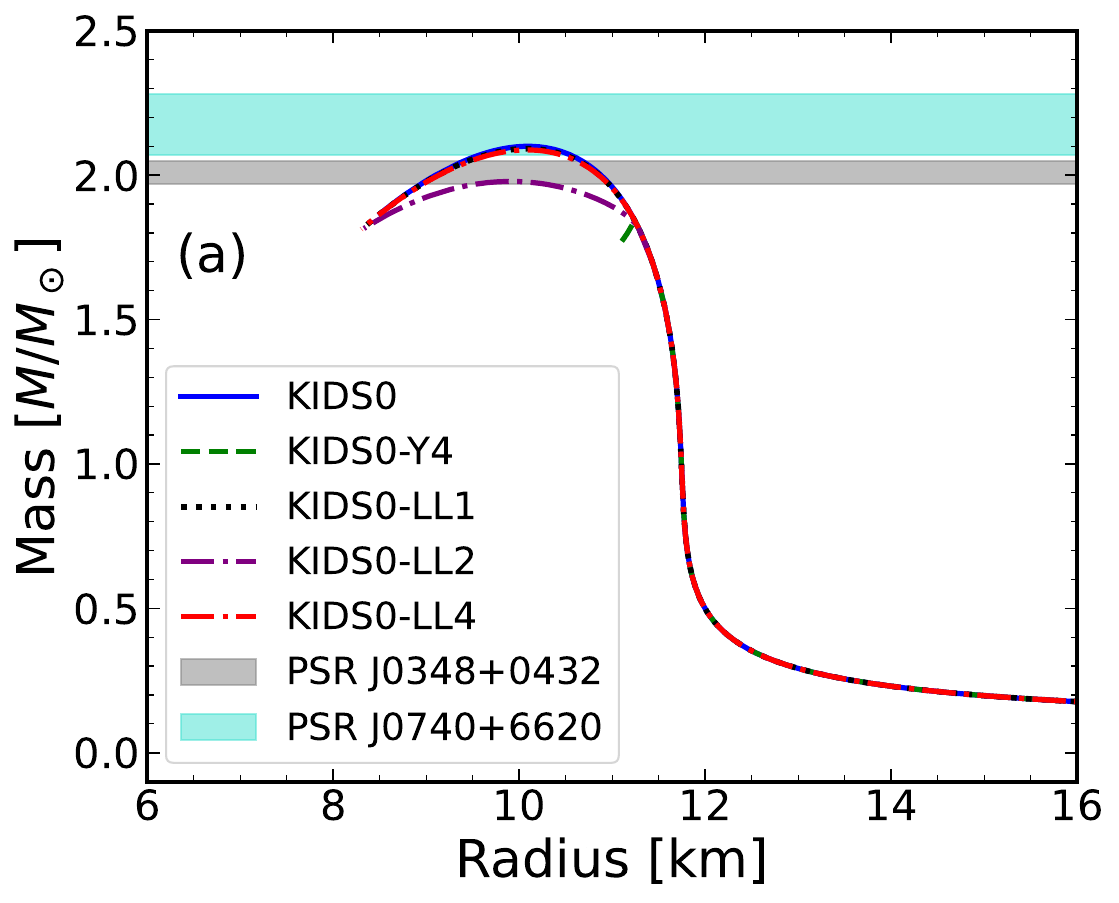}
    \includegraphics[width=0.45\textwidth]{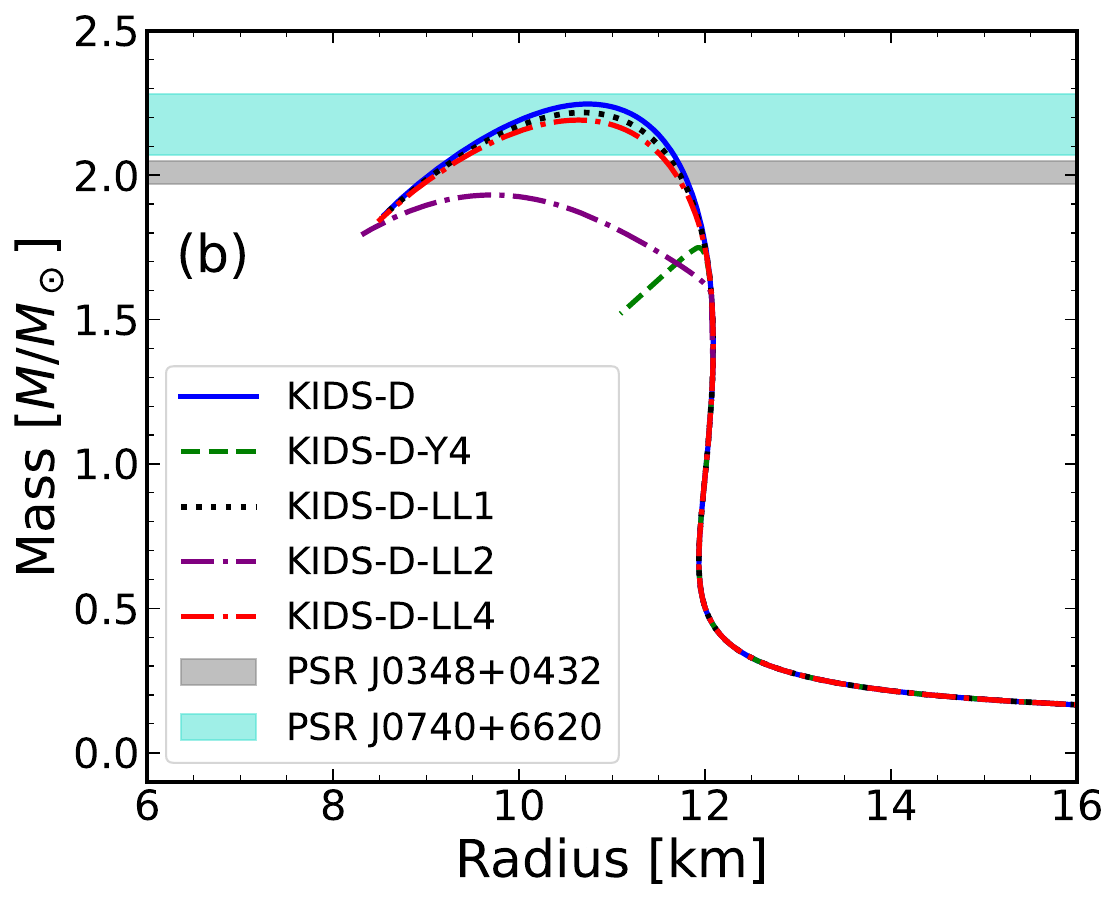}
    \end{center}
    \caption{Neutron star mass-radius relation for the (a) KIDS0 and (b) KIDS-D models.
    Models are the same as those in Fig.~\ref{fig3}.
    Data of PSR J0348+0432 and PSR J0740+6620 are from \cite{Antoniadis:2013pzd} and \cite{NANOGrav:2019jur}, respectively.}
    \label{fig4}
\end{figure*}

Figure \ref{fig3} depicts the pressure as a function baryon density
with $NN$ interactions only (blue solid line), with $NN$ and $N\Lambda$ interactions (green dashed line),
and with LL interactions for the KIDS0 (left) and KIDS-D (right) models. 
Addition of the $N\Lambda$ interaction leads to enormous softening of the EoS.
It is caused by explosively increasing population of the $\Lambda$ hyperon, which reduces the pressure exerted by the Pauli blocking of the nucleon.
However, inclusion of the $\Lambda\Lambda$ interaction changes the situation dramatically again to stiffer EoS.

For the KIDS0-LL1 and KIDS0-LL4 models, fraction of $\Lambda$ does not exceed 6~\% even at the center of the neutron star,
so their $\Lambda\Lambda$ effect to the EoS is marginal.
For the KIDS-D model, since $\Lambda$ hyperons are created at relatively low densities and $\Lambda$ fraction is larger than
the KIDS0 model, role of the $\Lambda$ hyperon becomes more evident.
However, it also softens the EoS only slightly, and consequently the resultant EoS is similar to the EoS of nucleon matter.

On the other hand, abrupt increase of the $\Lambda$ fraction in the KIDS0-LL2 and KIDS-D-LL2 models
causes unstable region in which $\partial P /\partial \rho <0$.
The interval of the unstable region is very short, and the EoS recovers to the normal behavior quickly.
But, since the $\Lambda$ fraction is more than 30~\% for the LL2 parameters, it gives rise to a substantial softening of the EoS.
Softening is more prominent in the KIDS-D-LL2 model because $\Lambda$ hyperons are created at low densities
and its fraction is more than 40~\% in the core of the neutron star.

In brief, the portion of $\Lambda$ hyperons in Fig. \ref{fig2} plays a critical role on the EoS in Fig. \ref{fig3}. The smaller $\Lambda$ hyperons by the repulsive $\Lambda\Lambda$ interaction produced in the high density region leads to stiffer EoS in the neutron star.

\subsection{Mass-Radius relation}


\begin{figure*}
    \begin{center}
    \includegraphics[width=0.45\textwidth]{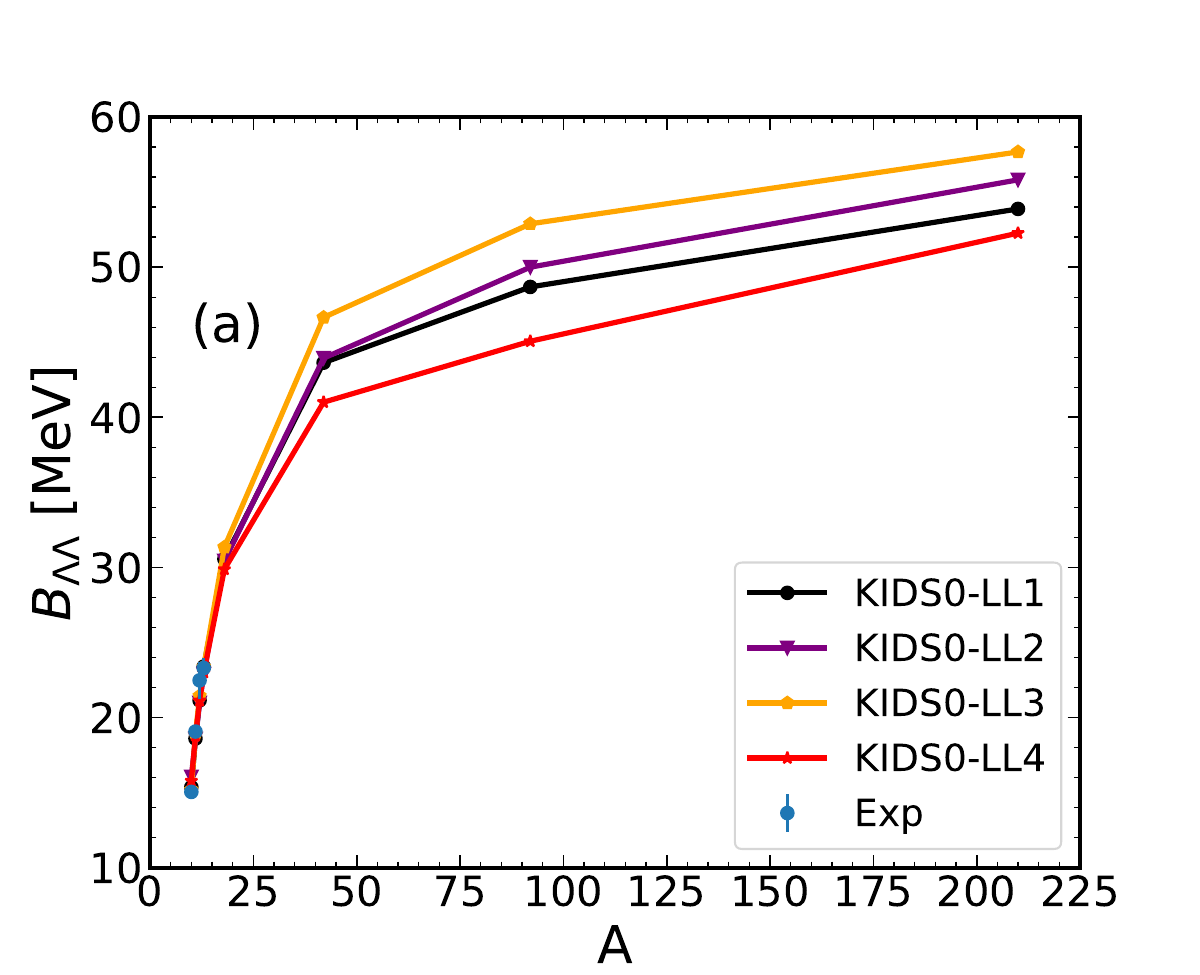}
    \includegraphics[width=0.45\textwidth]{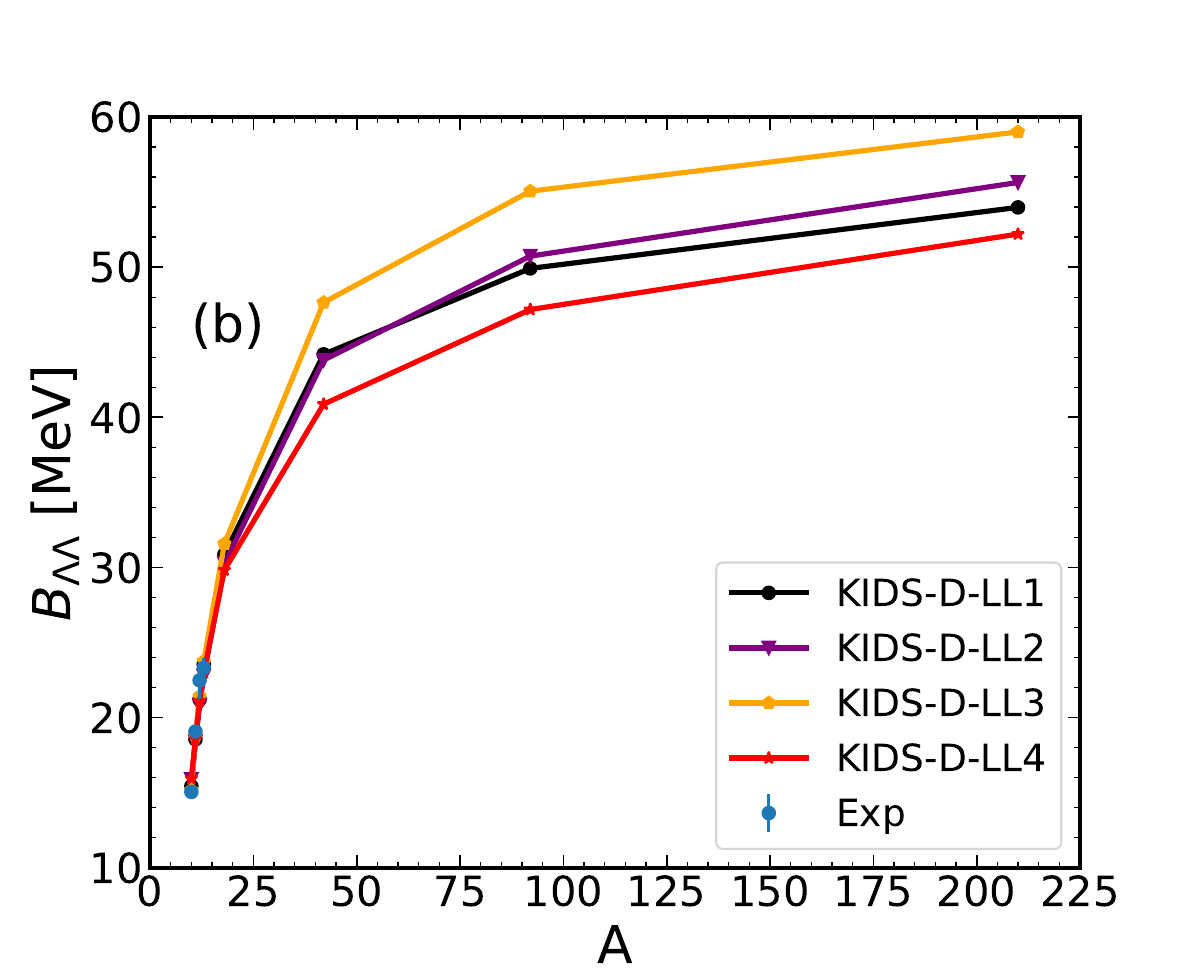}
    \end{center}
    \caption{Predictions of the binding energies for the heavy double-$\Lambda$ nuclei $^{18}_{\Lambda\Lambda}$O, 
    $^{42}_{\Lambda\Lambda}$Ca, $^{92}_{\Lambda\Lambda}$Zr, and $^{210}_{\Lambda\Lambda}$Pb for the (a) KIDS0-LL and (b) KIDS-D-LL models. }
    \label{fig5}
\end{figure*}

Bulk properties of the neutron stars such as mass, radius, tidal deformability and moment of inertia are direct consequences of the EoS.
Recent astronomical observations such as NICER, GW and LMXB provide unprecedented constraints on the mass and radius of the neutron star.
Combination of the data suggests the radius of a canonical mass ($1.4 M_\odot$) star in the range $R_{1.4} = 11.8$--$13.1$ km \cite{Raaijmakers:2019qny}.
One important issue is whether the exotic phases such as the mixture of hyperons or transition to deconfined quark matter will
affect the properties of the $1.4 M_\odot$ star.

Results of the mass-radius relation are shown in Fig.~\ref{fig4}.
KIDS0 and KIDS-D models (no $\Lambda$) in which there are only nucleons satisfy the radius constraint of the canonical star $R_{1.4} = 11.8$--$13.1$ km.
Without exotic phases, they also produce the maximum mass consistent with the large mass observation ($\geq 2 M_\odot$).
Hyperon puzzle is that the maximum mass is not as large as $2\;M_\odot$ when hyperons are created in the neutron star.
In the KIDS0-Y4 and KIDS-D-Y4 models, soon after the hyperons are created, the EoS becomes so soft that it cannot resist
the compression by the gravitation.
As a result, maximum mass does not reach $2\;M_\odot$.
However, the effect of hyperons emerges at masses larger than $1.5\;M_\odot$, so it has no effect to the 
properties of $1.4\;M_\odot$ stars.  

When the $\Lambda\Lambda$ interaction is taken into account,
the issue of the hyperon puzzle disappears, or becomes moderate.
For the KIDS0-LL1 and KIDS0-LL4 models, since the EoS is almost the same with the EoS of nucleon matter,
mass-radius relations are also similar to the result of KIDS0 model.
For the KIDS-D-LL1 and KIDS-D-LL4 models, softening of the EoS due to the $\Lambda$ hyperon is more remarkable than the KIDS0-LL1 
and KIDS0-LL4 models,
but the effect is not significant that the maximum masses are within the range of large mass observations.
Interesting is the behavior of the LL2 model.
Sudden increase of the $\Lambda$ fraction makes the mass-radius relation deviate from the curve of nucleon matter,
and increase of the mass is significantly slowed down.
Nevertheless, the mass keeps increasing, and it reaches $2.0\;M_\odot$ for KIDS0-LL2 and $1.95\;M_\odot$ for KIDS-D-LL2.
It is evident that the repulsive $\Lambda\Lambda$ interaction plays a critical role in the matter composition and the EoS of the neutron star core.
$\Lambda\Lambda$ interactions we determine in this work still contain substantial uncertainty, but they commonly
contribute to stiffening the EoS, and give more positive signal to the resolution of the hyperon puzzle.
Here we note that similar conclusion of the importance of the $\Lambda\Lambda$ interaction has been done by studying the odd-state $\Lambda\Lambda$ interaction by p-wave interaction using a variational principle by one of present coauthors~\cite{Togashi:2016fky}.

In this work, we consider the $\Lambda$ hyperon only, but it is well known that $\Sigma$ and $\Xi$ hyperons can contribute to EoS.
In general, they are created at densities higher than the $\Lambda$ hyperon because
their masses are larger, and their interactions are less attractive than the $\Lambda$ hyperon.
Therefore, even though the $\Sigma$ and $\Xi$ hyperons are considered in the neutron star EoS,
they are not likely to affect the properties of $1.4\;M_\odot$ stars.
Maximum mass could be more sensitive to inclusion of them.
Therefore, it is premature to draw a definite conclusion about the hyperon puzzle from the present result.
However, it is certain that the role of $\Lambda\Lambda$ interaction is important and critical in the EoS
of neutron stars at high densities.

%
\subsection{Heavy double $\Lambda$ nuclei}

We have seen that the sign and magnitude of the parameters $a_1$, $a_2$ and $a_3$ play a crucial role in the EoS of the neutron star.
In particular, the difference between LL1 and LL4 are not discernible in the properties of neutron star (see Fig.~\ref{fig3} and Fig.~\ref{fig4}).
Laboratory experiments are certainly helpful in reducing the uncertainties of the parameters.
Experimental data for the $\Lambda\Lambda$ nuclei are limited to light nuclei only,
but we can observe a specific trend for the $\Lambda\Lambda$ interaction.
Table~\ref{tab2} shows that the LL1 interactions tend to give overbinding for the $_{\Lambda\Lambda}$B nuclei.
It is questionable whether such a trend will persist in heavy nuclei.
Figure~\ref{fig5} displays the result of theory for the binding energies of $\ce{^{18}_{\Lambda\Lambda}}$O, 
$\ce{^{42}_{\Lambda\Lambda}}$Ca, $\ce{^{92}_{\Lambda\Lambda}}$Zr, and $\ce{^{210}_{\Lambda\Lambda}}$Pb.
Left and right panels show the result of KIDS0 and KIDS-D, respectively.
Interestingly, for each $\Lambda\Lambda$ interaction, results are insensitive to the symmetry energy,
so the KIDS0 and KIDS-D models predict similar binding energies.
However, the dependence on the $\Lambda\Lambda$ interaction becomes transparent in each model.
In the KIDS-D model, for example,
binding energy differences between LL1 and LL4 interactions are 5, 8, 7, and 4 MeV for $^{18}_{\Lambda\Lambda}$O, 
$^{42}_{\Lambda\Lambda}$Ca, $^{92}_{\Lambda\Lambda}$Zr, and $^{210}_{\Lambda\Lambda}$Pb, respectively.
We suggest that the binding energy of $\ce{^{42}_{\Lambda\Lambda}Ca}$ and $\ce{^{92}_{\Lambda\Lambda}Zn}$ could indicate the importance of the many-body interactions of the $\Lambda\Lambda$ in the LL1 and LL4 models.
Magnitude of the difference is large enough to be distinguished easily in the experiment.
If sufficiently accurate measurements could be performed for the heavy nuclei,
it will provide valuable constraints to the nature of three-body forces in the $\Lambda\Lambda$ interaction.

\section{Summary}

We implemented $\Lambda\Lambda$ interaction in the previous KIDS model which included only $\Lambda$ interaction. 
In the present work, we have implemented p-wave $\Lambda\Lambda$ interaction and taken into account the density-dependent many-body interactions including $\Lambda\Lambda$ interaction 
by considering the updated $\Lambda\Lambda$ nuclei data. 
Particularly, the roles of p-wave interaction as well as the many-body interaction which has $a_3$ coefficient in the term proportional to the density of $\rho_{\Lambda}^2 \rho_N$ term are investigated in detail with the momentum-dependent $\Lambda\Lambda$ interactions.

The many-body effects turn out to be smaller than the experimental errors in the binding energies of finite nuclei (see Fig.~\ref{fig1}). 
But, in nuclear matter, the $\Lambda\Lambda$ interactions give rise to the strong reduction of the $\Lambda$ fractions produced by the $\Lambda N$ interaction beyond the normal nuclear density region because of the repulsive momentum-dependent $\Lambda\Lambda$ ($\rho\tau$) interaction (see Fig.~\ref{fig2}) coming mainly from the p-wave interaction.
Consequently, the equation of the state (EoS) in Fig.~\ref{fig3} becomes stiffer, which enables us to explain the massive neutron star ranging to about $2.0\;M_\odot$, similarly to the results only by nucleons (see Fig.~\ref{fig4}).

But the problems whether the many-body interaction is repulsive (LL2 and LL4 models) or attractive (LL1 models) in finite nuclei and how the properties of the interaction are changed in nuclear matter are still unclear because both models provide almost same pressure in Fig.~\ref{fig3} and the mass-radius relation of neutron star data in Fig.~\ref{fig4}, although the LL2 models do not reach to $2.0\;M_\odot$. Therefore, we illustrated the binding energies of heavier $\Lambda\Lambda$ nuclei and suggested that $\ce{^42_{\Lambda\Lambda}}$Ca and $\ce{^92_{\Lambda\Lambda}}$Zn could discern the many-body interaction (see Fig. 5).

Decreased portion of $\Lambda$ hyperons by the repulsive momentum-dependent interaction in neutron star turns out to be critical for explaining the EoS and the mass-radius relation of neutron stars. More constraints in those models could be possible in near future if we compare our results to 
more data from the binding energies of heavier $\Lambda\Lambda$ nuclei as well as the data from astrophysical physics sides. 
We leave it as the next project.

\section*{Acknowledgments}
The work of SC is supported by the Institute for Basic Science (IBS-R031-D1).
The work of EH was supported by JSPS KAKENHI Grant Numbers JP18H05407 and JP20H00155.
The work of CHH was supported by the National Research Foundation of Korea (NRF) grant
funded by the Korea government (No. 2018R1A5A1025563 and No. 2023R1A2C1003177).
The work of MKC was supported by the National Research Foundation of Korea (NRF) grant
funded by the Korea government (No.2021R1A6A1A03043957 and No. 2020R1A2C3006177).

\bibliography{ref.bib}

\end{document}